\documentclass[prb,twocolumn,showpacs,preprintnumbers,amsmath,amssymb,citeautoscript,floatfix]{revtex4}

\usepackage{graphicx}
\usepackage{dcolumn}
\usepackage{bm}
\usepackage{color}
\usepackage{amsmath}
\usepackage{multirow}
\usepackage{mathrsfs} 

\begin{document}

\title{\boldmath Node-like excitations in superconducting PbMo$_6$S$_8$ probed by
scanning tunneling spectroscopy}

\author{C. Dubois}
 \email{cedric.dubois@physics.unige.ch}
\author{A. P. Petrovi\'{c}}
\author{G. Santi}
\author{C. Berthod}
\author{A. A. Manuel}
\author{M. Decroux}
\author{\O. Fischer}
\affiliation{ DPMC-MaNEP, Universit\'e de Gen\`eve, Quai Ernest-Ansermet
24, 1211 Gen\`eve 4, Switzerland}
\author{M. Potel}
\author{R. Chevrel}
\affiliation{Sciences Chimiques de Rennes, CSM UMR CNRS 6226,
Universit\'e de Rennes 1, Avenue du
G\'en\'eral Leclerc, 35042 Rennes Cedex, France}

\date{\today}

\begin{abstract}
We present the first scanning tunneling spectroscopy study on the Chevrel
phase PbMo$_6$S$_8$, an extreme type II superconductor with a coherence
length only slightly larger than in high-$T_c$ cuprates. Tunneling spectra
measured on atomically flat terraces are spatially homogeneous and show
well-defined coherence peaks. The low-energy spectral weight, the zero bias
conductance and the temperature dependence of the gap are incompatible with
a conventional isotropic $s$-wave interpretation, revealing the presence of
low-energy excitations in the superconducting state. We show that our data
are consistent with the presence of nodes in the superconducting gap.
\end{abstract}


\maketitle

The Chevrel phase family of superconductors attracted great attention in
the 1970s and '80s since several of its members possessed unprecedentedly
high upper critical fields while others displayed astonishing properties
resulting from the coexistence of superconductivity and magnetism
\cite{Fischer-82}.  Interest in these compounds faded somewhat with the
advent of high temperature superconductors (HTS). However, in the quest to
understand the pairing mechanism in HTS, research has recently diversified
into other strongly correlated electron systems which exhibit unusual
superconductivity. For instance, heavy
fermions~\cite{Steglich-79,Radovan-03},
borocarbides~\cite{Cava-94,DeWilde-97}, organics~\cite{Jerome-80,Arai-01}
and ruthenates~\cite{Maeno-94,Upward-02} have already provided vital
insight into the fundamental questions of gap symmetry and the interplay
between superconductivity and magnetism.  In this context, the properties
of the Chevrel phases render them particularly relevant and interesting.
Among these compounds, PbMo$_6$S$_8$ is one of the most remarkable with its
high critical temperature of 14~K and a strikingly high upper critical
field of about 60~T, corresponding to a very short coherence length $\xi
\approx 23$ {\AA} (estimated from the critical field \cite{Fischer-78}
using Ginzburg-Landau theory). This value is comparable to those found in
the HTS (e.g. 11, 20, 30~{\AA} for Bi$_2$Sr$_2$CaCu$_2$O$_{8+\delta}$,
YBa$_2$C$_3$O$_{7-\delta}$, La$_{2-x}$Sr$_x$CuO$_4$ respectively
\cite{Poole-95}). With a $\kappa=\lambda/\xi$ above 100, PbMo$_6$S$_8$ is
clearly an extreme type II superconductor.

Scanning Tunneling Spectroscopy (STS) today plays a central role in the
study of novel superconductors at an atomic scale. In particular, STS
investigations have revealed significant differences between classical BCS
and high temperature superconductors~\cite{Fischer-06}, notably the very
large gaps in the spectra consistent with $d$-wave symmetry and the spatial
dependence of the electronic structure in the vortex cores which differ
radically from what is observed in conventional
superconductors~\cite{Hess-90,Renner-91}.  One possible contribution to
these differences is the very small size of the Cooper pairs in HTS,
resulting in a deviation from mean field theory.  Compared with HTS,
PbMo$_6$S$_8$ is stoichiometric, its structure 3D and, unlike other Chevrel
phases, is not known to display competition between superconductivity and
magnetism. We consider it to be a bridge between classical type-II
superconductors and HTS.  When the Chevrel phases were first studied, it
was assumed during most investigations that they were classical $s$-wave
superconductors~\cite{Fischer-82} similar to other type II superconductors
known at that time.  However, the very small Cooper pair size raises
questions over the microscopic nature of the superconducting state in
PbMo$_6$S$_8$ and, more generally, the role of short coherence length in
unconventional superconductivity.

In this paper we present the first STS study of PbMo$_6$S$_8$ single crystals.
We show tunneling spectra measured on atomically flat cleaved surfaces at
different temperatures.  The central result reported here is that the
spectra obtained reproducibly on different crystals are clearly
incompatible with the classical $s$-wave model and reveal the presence of
low-energy excitations between the superconducting coherence peaks.

The high-quality PbMo$_6$S$_8$ single crystals were prepared by both
chemical vapor transport in silica tubes and high temperature melt
techniques in sealed Mo crucibles~\cite{Chevrel-crystal}. They display a
sharp superconducting transition at $T_{c} = 14$~K with width $\Delta
T_{c}=0.3$~K determined by ac susceptibility measurements. PbMo$_6$S$_8$
crystallizes in a ``quasi-cubic'' rhombohedral lattice, space group
$R\overline{3}$~\cite{Fischer-82}, with the three-fold axis along the [111]
direction. We obtained flat (001) surfaces by cleaving the crystals using a
blade under an optical microscope.  The STS measurements were carried out
using a home-built scanning tunneling microscope, featuring an
XY-stage~\cite{Dubois-06} to target the micrometer-sized crystals. We use
electrochemically etched Ir tips and measure the differential conductivity
using a standard lock-in technique.

\begin{figure}[tb]
\includegraphics [width=\linewidth,clip] {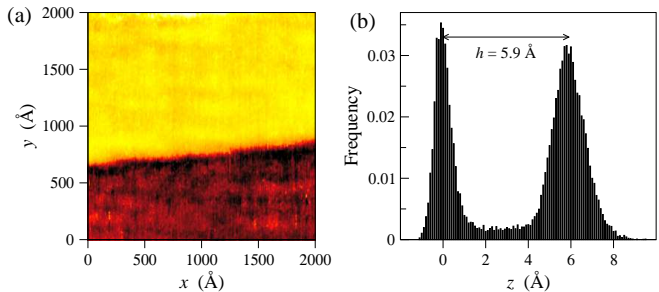}
\caption{\label{f:topo}Topographic characterization. (a) Large-scale
topography ($T=1.8$~K, $R_{t}= 50~\text{M}\Omega$). (b) Corresponding
height histogram showing a unit step of height $5.9 \pm 1.1$~\AA.}
\end{figure}

The surface topography imaged at 1.8~K with the tip perpendicular to the
(001) plane reveals large flat terraces separated by long parallel steps.
Fig. \ref{f:topo} shows a typical step of height $5.9 \pm 1.1$~\AA. This
corresponds to the distance between the Pb planes which are the natural
cleavage planes of PbMo$_6$S$_8$.  The surfaces investigated were
atomically flat with a RMS roughness under 0.7~\AA.

\begin{figure}[tb]
\includegraphics [width=\linewidth,clip] {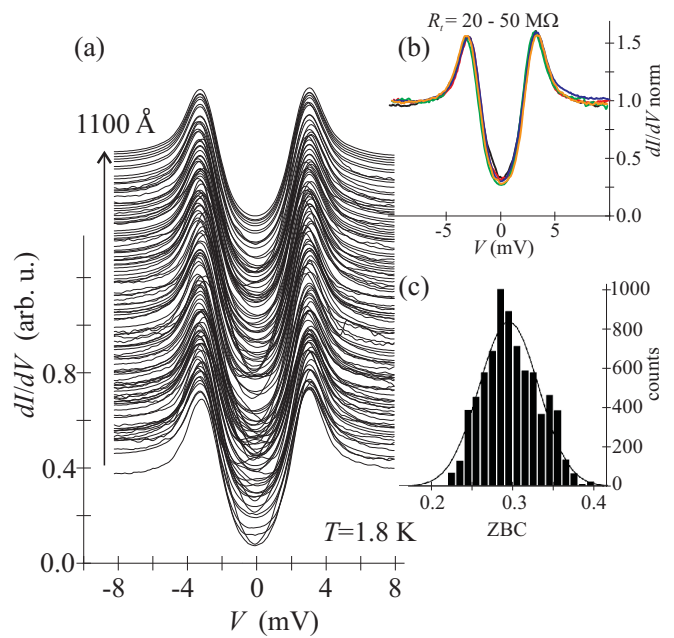}
\caption{\label{f:spectrobroken} (a) Spectroscopic trace along a 1100~{\AA}
path with one spectrum every 8.8~\AA. The spectra are offset for clarity
($T=1.8$~K, $R_{t}=25~\text{M}\Omega$). (b) Normalized conductance spectra
obtained at various $R_{t}$ from 20 to 50~M$\Omega$.  (c) Histogram of the
zero bias conductance in $\sim 7500$ spectra normalized to the conductance
at 8 mV. The standard deviation is 0.04. }
\end{figure}

Homogeneous spectra have been obtained at a large number of positions on
different crystals along lines (Fig.~\ref{f:spectrobroken}a) and on maps
extending over a range of several hundred {\AA}ngstr\"{o}ms.  We have
verified that the spectra obtained by varying the tunnel resistance $R_{t}$
all collapse onto a single curve (Fig.~\ref{f:spectrobroken}b), which
confirms true vacuum tunneling conditions.  It is also important to mention
that our measurements have always shown the same spectroscopic signature
regardless of sample and cleavage. The spectra present well-developed and
homogeneous coherence peaks.  The zero bias conductance (ZBC) is
homogeneous with a value of $\sim 30\%$ of the high-energy conductance
(Fig.~\ref{f:spectrobroken}c).  This unusually large value, also reported
by early planar-junction experiments \cite{Poppe-1981}, could in principle
be due to an extrinsic influence such as surface impurities. However, the
high reproducibility of the ZBC leads us to believe it to be intrinsic to
this material. Furthermore, we stress that PbMo$_6$S$_8$ does not suffer
from surface contamination or ageing effects.

\begin{figure}[tb]
\centering
\begin{minipage}[t]{0.1\linewidth}
\flushright
\vspace{0pt}
(a)
\end{minipage}%
\begin{minipage}[t]{0.9\linewidth}
\flushleft
\vspace{0pt}
\includegraphics [width=\linewidth,origin=ct] {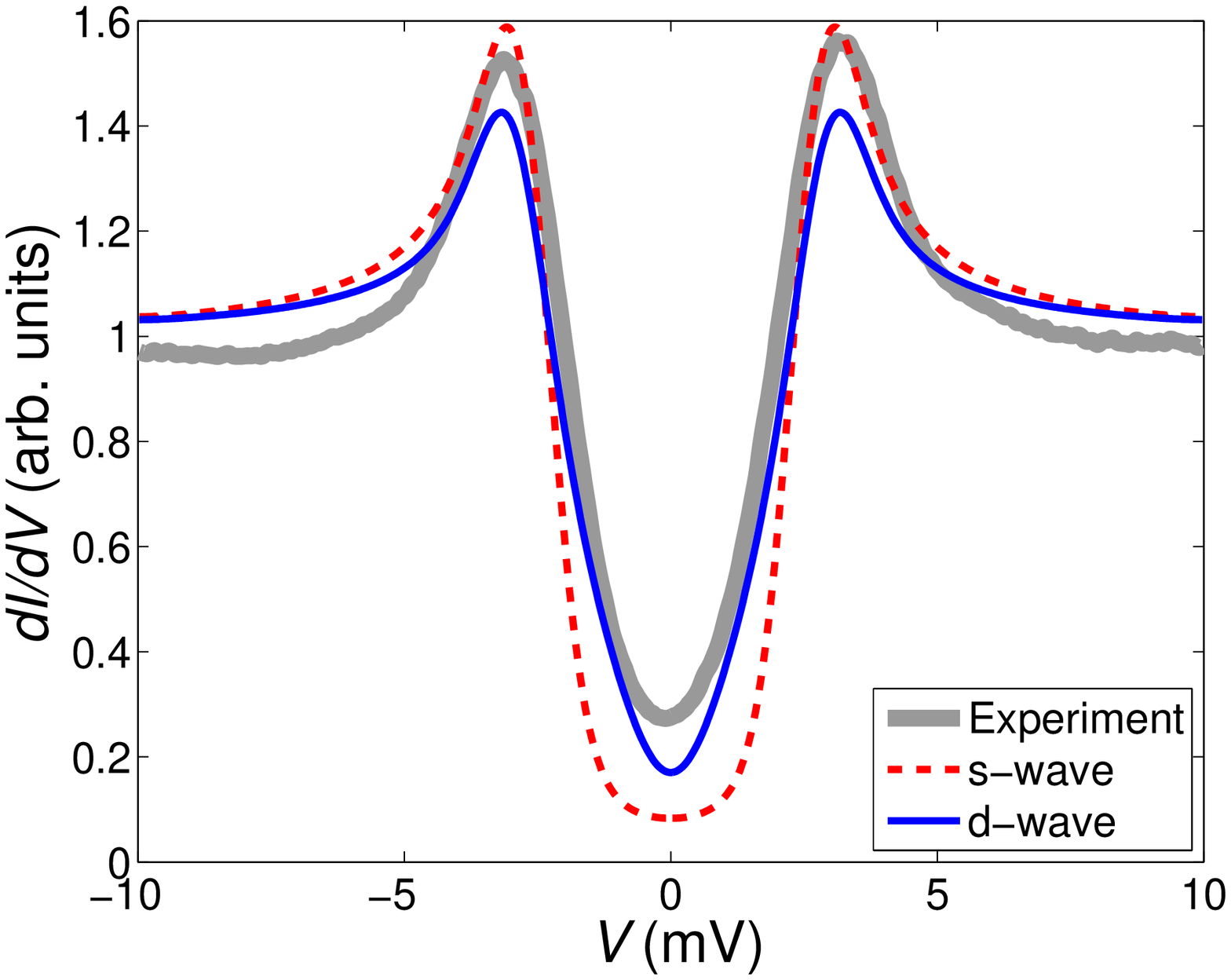}
\end{minipage}
\\
\begin{minipage}[t]{0.1\linewidth}
\flushright
\vspace{0pt}
(b)
\end{minipage}%
\begin{minipage}[t]{0.9\linewidth}
\flushleft
\vspace{0pt}
\includegraphics [width=\linewidth] {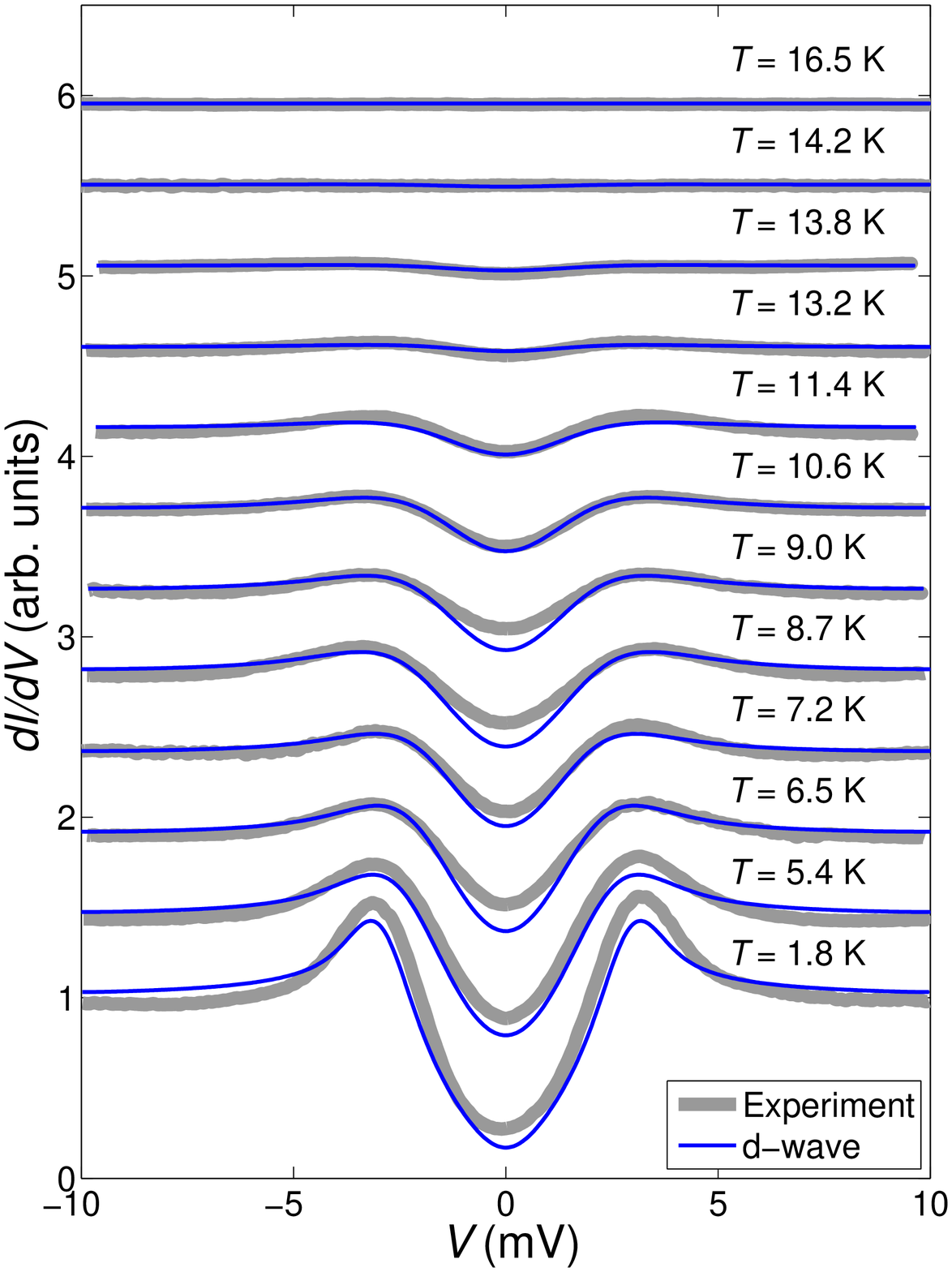}
\end{minipage}
\caption{\label{f:Fitdwave} (a) Experimental $dI/dV$ spectrum at 1.8~K
(thick line) compared to the $s$-wave and $d$-wave fits described in the
text (dashed and solid lines respectively).  (b) $dI/dV$ spectra obtained
at various temperatures (thick lines) and $d$-wave fits (thin lines). Each
experimental spectrum is normalized so as to ensure state conservation within
the measurement energy range and offset by 0.45 with respect to the
previous one for clarity.  }
\end{figure}

The shape of the spectra reveals the presence of low-energy excitations
which cannot be explained by thermal broadening and is reminiscent of that
observed in HTS (Fig.~\ref{f:Fitdwave}).  We have therefore performed fits
to our experimental spectra using a single-band model and different
singlet-gap symmetry scenarii.  STS is a direct probe of the local
quasiparticle DOS and is therefore not sensitive to the phase of the order
parameter. Hence it cannot determine whether a sign change occurs, but it
can reveal the presence of regions where the gap function is zero (nodes)
or nearly zero (node-like). Using the BCS Green's function and including
lifetime effects due to impurity scattering, the quasiparticle
density-of-states (DOS) reads:
\begin{equation}
N(\omega) = -\frac{1}{\pi} \sum_{\bm{k}} \mathrm{Im} \left[ 
\frac{\omega + \xi_{\bm{k}} + i \Gamma}
{(\omega + i
\Gamma)^2 - \xi_{\bm{k}}^2 - |\Delta_{\bm{k}}|^2} \right]
\end{equation}
where $\xi_{\bm{k}} = \epsilon_{\bm{k}}-\epsilon_{\text{F}}$ are the band
energies relative to the Fermi level, $\Delta_{\bm{k}}$ the superconducting
gap and $\Gamma$ the scattering rate.  We have calculated the DOS assuming
a spherically symmetric Fermi surface; the relevance of the real band
structure will be discussed below.

\begin{table}[b!]
\caption{\label{t:gap} Lowest-order representations compatible with
rhombohedral symmetry for the gap $\Delta_{\bm{k}} = \Delta_0
\psi(\bm{k})$. $\bm{\hat{z}}$ is chosen along the 3-fold axis.
}
\renewcommand{\arraystretch}{1.25}
\begin{ruledtabular}
\begin{tabular}{cccc}
  & Pairing & $\bm{k}$-dependence & Relation to  \\
  & symmetry \cite{dwavenote}& of the gap, $\psi(\bm{k})$ & point-group symmetry \\
  \hline
  $\mathscr{R}_1$ & $s$ & 1 & Preserving (isotropic) \\
  \hline
  $\mathscr{R}_2$ & $d$ & $(k_x \pm ik_y)^2$ & \multirow{2}*{Breaking} \\
  $\mathscr{R}_3$ & $d$ & $k_z(k_x \pm ik_y)$ &  \\
  \hline
  $\mathscr{R}_4$ & $d$ & $(k_x^2+k_y^2)$ & \multirow{2}*{Preserving} \\
  $\mathscr{R}_5$ & $d$ & $\frac{1}{2}(3k_z^2-1)$ &  \\
\end{tabular}
\end{ruledtabular}
\end{table}

The gap representations compatible with the rhombohedral symmetry are very
similar to those for the hexagonal group in Ref.~~\onlinecite{Sigrist-91}.
We shall focus on the lowest-order gaps ($s$ and $d$) which are listed in
Table~\ref{t:gap}.  Representations $\mathscr{R}_2$ and $\mathscr{R}_4$
give identical DOS as they only differ by a phase factor. Furthermore,
representation $\mathscr{R}_3$, although behaving differently close to the
nodes, gives a DOS which is indistinguishable from that of $\mathscr{R}_2$
and $\mathscr{R}_4$ once integrated over the Fermi surface. We thus group
these three representations under the collective label ``$d$-wave''
\cite{dwavenote}.  In contrast, the typical DOS from the last $d$-wave
representation, $\mathscr{R}_5$, shows greatly reduced peaks (similar to
higher-order $g$-wave cases not shown in Table~\ref{t:gap}) which are
totally incompatible with the observed spectra and will therefore not be
further discussed here.

The average experimental spectrum at 1.8~K is shown in
Fig.~\ref{f:Fitdwave}a together with the fits. The calculated spectra take
into account temperature smearing and a Gaussian broadening with a
half-width at half-maximum of 0.4~mV. This value corresponds to the lock-in
amplitude and sets the maximum experimental resolution. The fits are
defined to optimally reproduce the peak energy positions and widths. The
isotropic $s$-wave for $\Delta_0=2.6$ meV and $\Gamma=0.2$ meV clearly
fails to describe the spectrum between the coherence peaks. In direct
contrast, the $d$-wave cases for $\Delta_0=2.95$ meV and $\Gamma=0$ meV
provide a much better fit in the low-energy region which is the most
sensitive to changes in the gap symmetry.  Attempting to fit the ZBC with a
$s$-wave symmetry by raising $\Gamma$ or introducing a normal residual DOS
\cite{residualdos} fails either due to a dramatic coherence peak reduction
or a poor description of the spectrum at low energies.  Furthermore, a
mixed-symmetry gap function (e.g. $\mathscr{R}_1 + \mathscr{R}_4$) always
results in a deterioration of the fit in the low-energy region.  We
emphasize that within the $d$-wave scenario, the measured ZBC is a natural
consequence of the low-lying excitations and thermal smearing. Note that
the experimental coherence peaks are larger than those from $d$-wave
fits. One possible explanation for this could be a spectral weight transfer
from high energies to the peaks due to strong coupling effects
\cite{Schrieffer-1963,McMillan-1965}.

\begin{figure}[tb]
\centering
\includegraphics [width=0.9\linewidth] {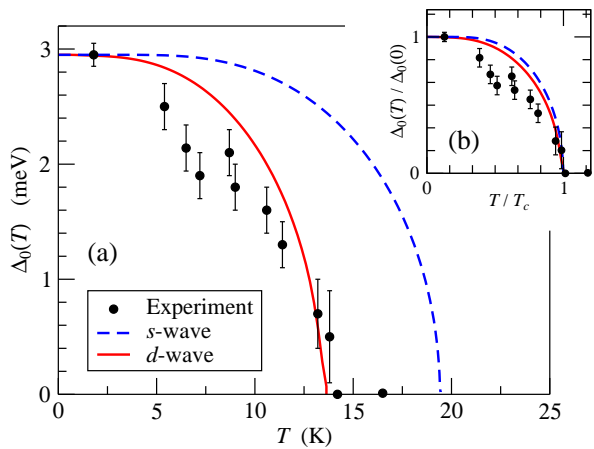}
\caption{\label{f:T}
(a) Temperature dependence of the superconducting gap from the spectra in
Fig.~\ref{f:Fitdwave}b (circles) compared to the $s$ and $d$-wave cases from
solving the BCS gap equation for $\Delta_0(0)=2.95$~meV 
(dashed and solid curves respectively). (b) Same data with gaps rescaled by
$\Delta_0(0)$ and temperatures rescaled by the respective $T_c$ values.}
\end{figure}

Fig.~\ref{f:Fitdwave}b shows spectra taken at different temperatures with
their corresponding $d$-wave fits. We first observe that superconductivity
measured at the surface is suppressed at exactly the bulk $T_{c}=14$~K.
This confirms that our measurements probe the bulk properties of
PbMo$_6$S$_8$. The $d$-wave models fit well at all temperatures, in
contrast to the $s$-wave which consistently fails to reproduce the spectral
shape at low energy.  The gap amplitudes $\Delta_0(T)$ from the fits to our
experimental spectra are plotted as a function of temperature in
Fig.~\ref{f:T}, together with the solutions of the BCS gap equation for
both the $s$-wave and $d$-wave symmetries.  These were calculated by
selecting the simplest separable coupling interaction preserving the gap
symmetry, i.e. $V_{\bm{k},\bm{k'}} = V_0 \psi(\bm{k}) \psi^*(\bm{k'})$, and
determining $V_0$ so as to fix the zero-temperature gap at our experimental
value (2.95~meV).  The $s$-wave solution gives a $T_c$ of 19.4~K, as
expected from the well-known $2\Delta_0/k_{\text{B}}T_c$ ratio of 3.52 but
clearly at variance with our experiment. On the other hand, the $T_c$ from
the 3D $d$-wave model is 13.6~K, giving a ratio of 5.0 in agreement with
our experimental value of $4.9 \pm 0.3$ (Fig.~\ref{f:T}a).  These results
are also shown renormalized by their respective $T_c$ values in
Fig.~\ref{f:T}b, highlighting the facts that the presence of nodes affects
the shape of the curve (and not only the $2\Delta_0/k_{\text{B}}T_c$ ratio)
and the experimental gap closes faster with increasing temperature than the
theoretical curves.

The large low-temperature gap is also consistent with the high upper
critical field: using the Fermi velocity from Ref.~~\onlinecite{Woollam-79},
$v_{\text{F}} \approx 4\times 10^{4}$ ms$^{-1}$, we obtain a rough estimate
of the coherence length $\xi=\frac{\hbar v_{\text{F}}}{\pi \Delta_0}
\approx 28~$\AA, in line with the value derived earlier from $H_{c2}$
\cite{Fischer-78}. This very small coherence length and the large coupling
ratio of 4.9 further confirm the similarity with the high-$T_c$ materials,
which has already been emphasized by \citet{Uemura-91}.

Let us now turn to the effect of the true band structure of this material.
Our FLAPW \cite{wien2k} calculations, in agreement with previous
calculations \cite{Fischer-82}, reveal that two Mo-$d$ bands (of width 0.7
eV) cross the Fermi level, both contributing roughly equally to the total
DOS at $\epsilon_{\text{F}}$. This justifies a single-band approach in
calculating the superconducting properties. The normal state DOS appears
constant at the energy scale of our measurements.  However, the calculated
Fermi surface (FS) is not spherically symmetric and exhibits significant
nesting at $\bm{Q} = 0.62$ and 1.08 $(2\pi/a)$ along the [111]
direction. Most strikingly, one of the two FS sheets displays two lobes
above and below the basal plane. This could be of importance in the case of
highly anisotropic superconducting gaps.  In effect, the presence of two
bands and the FS topology influence the gap distribution and could thus
explain the slight disagreement for the peak heights and ZBC between the
$d$-wave models and the observed spectra as well as the deviations from the
calculated temperature dependence.  The details of the Fermi surface could
also influence the spectra through the tunneling matrix element.  However,
since we tunnel from a direction $54^\circ$ off the main (3-fold) symmetry
axis, the STM tip effectively probes all relevant parts of the Fermi
surface.  Any further development of the model presented here would demand
the knowledge of the radial dependence of the gap function, itself
requiring a complete description of the pairing interaction.

In summary, we have measured the low-temperature characteristics of the
quasiparticle DOS of superconducting PbMo$_6$S$_8$ for the first time using
scanning tunneling spectroscopy.  In light of these results, we conclude
that a highly anisotropic gap function is the likely explanation for the
low-energy excitations observed in our spectra.  In particular, we have
shown that our measurements are well described by 3D $d$-wave models
compatible with the rhombohedral symmetry of PbMo$_6$S$_8$.  Such $d$-wave
superconductivity in PbMo$_6$S$_8$ could result from the very short
coherence length which would favor the appearance of a repulsive component
in the coupling interaction.  The combination of possible $d$-wave
symmetry, a large $2\Delta_0/k_{\text{B}}T_c$ ratio and short coherence
length in PbMo$_6$S$_8$ is strongly reminiscent of the HTS and calls for
further investigation. We believe that vortex core spectroscopy, which has
already revealed important differences between classical superconductors
and HTS, would be the ideal tool to confirm the role played by an intrinsic
short coherence length in unconventional superconductivity.

The authors particularly wish to thank M. Sigrist and F. Marsiglio for
invaluable discussions. This work was supported by the Swiss National
Science Foundation through the NCCR MaNEP.


\begin{thebibliography}{25}
\bibitem[{dwa()}]{dwavenote}
\bibinfo{note}{By $d$-wave, we mean representations corresponding to an orbital
  moment of $\ell = 2$ in the spherical harmonics expansion, some of which do
  not display a sign change.}
\bibitem[{res()}]{residualdos}
\bibinfo{note}{The residual DOS could originate from a band with a different
  character remaining normal (although this is not supported by our electronic
  structure calculations) or anomalous electronic states at the surface, for
  example.}
\end{thebibliography}

\end{document}